\documentstyle[aps,prl,twocolumn]{revtex}
\input{psfig}

\def\bra#1{ \langle{#1}| }
\def\braket#1#2{ \langle{#1}|{#2}\rangle }
\def\opket#1#2 {  {#1} |{#2}\rangle }
\def\braopket#1#2#3{ \langle{#1}| {#2} |{#3}\rangle }

\def\d{{\rm d}}

\def\K2{{\cal K}}
\def\ket#1{ |{#1}\rangle }

\def\T{{\cal T}}

\def\x{{\bf x}}

\begin{document}
\bibliographystyle{unsrt}
\newcommand{\be}{\begin{equation}}
\newcommand{\ee}{\end{equation}}
\newcommand{\bea}{\begin{eqnarray}}
\newcommand{\eea}{\end{eqnarray}}

\draft
\twocolumn[\hsize\textwidth\columnwidth\hsize\csname@twocolumnfalse%
\endcsname

\title
{The Statistics of Chaotic Tunnelling}

\author{Stephen C. Creagh}
\address{Division of Theoretical Mechanics, School of Mathematical
Sciences, University of Nottingham, NG7~2RD, UK}
\author{Niall D. Whelan}
\address{Department of Physics and Astronomy, McMaster University,
Hamilton, Ontario, Canada L8S~4M1}

\date{\today}

\maketitle

\begin{abstract}
We discuss the statistics of tunnelling rates in the presence of
chaotic classical dynamics. This applies to resonance widths in
chaotic metastable wells and to tunnelling splittings in chaotic
symmetric double wells. The theory is based on using the properties of
a semiclassical tunnelling operator together with random matrix theory
arguments about wave function overlaps. The resulting distribution
depends on the stability of a specific tunnelling orbit and is
therefore not universal. However it does reduce to the universal
Porter-Thomas form as the orbit becomes very unstable. For some
choices of system parameters there are systematic deviations which we
explain in terms of scarring of certain real periodic orbits. The
theory is tested in a model symmetric double well problem and possible
experimental realisations are discussed.
\end{abstract}

]

Tunnelling is crucial in describing many physical phenomena, from
chemical and nuclear reactions to conductances in mesoscopic devices
and ionisation rates in atomic systems. When such systems are complex, 
it is natural to model tunnelling effects using random matrix theory
\cite{rmt}. We show here that when the underlying system is one of
clean chaotic dynamics, successful statistical modelling demands
explicit incorporation of nonuniversal, but simple, dynamical
information. We derive a distribution for the tunnelling rate which
depends on a single parameter, calculated from the stability
properties of the dominant tunnelling orbit.

The signatures of chaos in tunnelling rates have been receiving a
growing amount of attention, two important regimes having been
considered. The first is that the quantum state is initially localised
in a region where the dynamics is largely nonchaotic and one wants to
understand the tunnelling rate through chaotic regions of phase space
\cite{cat,pofy_cat,si}. The second, and the one we shall focus on, is
that virtually all of the energetically accessible phase space is
chaotic \cite{us1us2,us3} so that the quantum state is initially
localised in a chaotic region of phase space. These two situations are
different in many important ways. In particular, the statistical
distribution of the tunnelling rates in the first regime has power law
decays \cite{pofy_cat} whereas we show that the distribution in the
second regime has exponential decay. The result is a generalisation of
the Porter-Thomas distribution \cite{porter} used to model neutron and
proton resonances \cite{porter,brody,prot} and conductance peak
heights in quantum dots \cite{dot,dot_ex}.

It is shown in Refs.~\cite{us1us2} that the average tunnelling rate
from an energetically-connected region of phase space is determined 
by a complex orbit we will call the instanton. Fluctuations about this 
average appear to be pseudo-random in the chaotic case and are given 
by properties of the wavefunction in an area localised around a real 
extension of the instanton \cite{us3}. To characterise these 
fluctuations we define a rescaled tunnelling rate as follows. 
In the case of metastable wells the absolute tunnelling rate of a 
given state labelled by $n$ is measured by the resonance width, 
or inverse lifetime $\Gamma_n$. The corresponding normalised  
tunnelling rate is defined 
to be
\begin{equation}
y_n = {\Gamma_n / \bar{\Gamma}},
\end{equation}
where $\bar{\Gamma}(E,\hbar)=\langle\Gamma_n\rangle$ is a local
average computed for a given set of physical parameters.
$\bar{\Gamma}(E,\hbar)$ is a smooth, monotonic function of its
arguments and is given by an explicit formula in terms of the (purely
imaginary) action and stability of the instanton \cite{us1us2}. A
similar definition holds for splittings in double wells and in either
case $\langle y_n\rangle=1$ by construction.


Fluctuations in $y_n$ are calculated using a {\it tunnelling operator},
$\T$, which is constructed from the semiclassical Green's function and 
can be interpreted as transporting the wavefunction across the barrier.
Specifically,
\begin{equation}
	y_n \propto \braopket{n}{\T}{n},
\end{equation}
where $\ket{n}$ is the wavefunction (which may be calculated
while ignoring tunnelling effects) represented in a 
Hilbert space which quantises a surface of section. A
closed-form expression can be found by 
%
%
expanding classical actions in quadratic order around the instanton.  
Details can be found in
\cite{us3}, but for present purposes it is enough to know its
spectrum. For a two-dimensional system this is
$\{\lambda^k\left|\lambda\right|^{1/2},k=0,1\cdots\}$, where $\lambda$
is the inverse of the stability of the instanton orbit, is always less
than unity in magnitude and can easily be found using real dynamics in
the inverted potential. (The discussion is readily generalised to
higher dimension but we refrain from doing so for clarity.)
To derive statistical distributions for $y_n$, we will make
statistical assumptions about the state $\ket{n}$, but not about the
tunnelling operator. The resulting distribution depends parametrically
on $\lambda$ and is therefore system-specific and not universal.

We can always express $\T$ as a diagonal operator in its own
eigenbasis: $\sum_k\lambda^k\left|\lambda\right|^{1/2}\ket{k}\bra{k}$. 
We then have
\begin{equation} \label{yn}
y_n = a\sum_{k=0}^\infty\lambda^k|\braket{k}{n}|^2 = 
a\sum_{k=0}^\infty\lambda^k|x_k|^2
\end{equation}
where we denote $x_k=\braket{k}{n}$ and the prefactor $a=1-\lambda$
ensures that $\langle y\rangle=1$.  The states $\ket{n}$ are
normalised so that on average $|x_k|^2$ is unity. We now make the
statistical ansatz that the overlaps can be treated as Gaussian random
variables. This is the basis of almost all statistical treatments of
wave functions, going back to the seminal work of Porter and Thomas
\cite{porter}. In the event that there is a time reversal symmetry,
$x_k$ can be expressed as a single real number, leading to GOE
statistics. If there is no such symmetry then $x_k$ is complex and
will be described by two statistically independent quantities leading
to GUE statistics. 

We simplify the derivation by assuming GOE statistics; the
generalisation to GUE is simple and we give the final result for
both. We start by assuming that the $x_k$ are statistically
independent
%
%
and given by the joint distribution
$P(\x)\d\x=\prod_k\left[\exp(-x_k^2/2)/\sqrt{2\pi}\right]\d x_k$ where
$\x=\{x_k\}$. We then note that
\bea
P(y;\lambda) & = & \int  \d\x\; P(\x)\;
\delta \Bigl(y-a\sum_{k=0}^\infty\lambda^kx_k^2\Bigr).
\eea
We use the identity $\delta(z)=\int \d t\exp(itz)/2\pi$ in the above
expression and observe that each $x_k$ involves a simple Gaussian
integral. The final result (and generalising to the GUE case) is
\begin{equation}\label{pofy} 
P(y;\lambda) = {1\over 2\pi}\int \d t\; e^{ity} \prod_{k=0}^\infty
\Bigl(1+{2i\over\beta} a\lambda^kt\Bigr)^{-\beta/2},
\end{equation}
where $\beta=1$ and $2$ for GOE and GUE respectively.  The product
above converges rapidly provided $\lambda$ is not too close to unity
so the formula can easily be used to calculate $P(y;\lambda)$ in
practice.

This result has a simple interpretation if we think of each eigenstate
of $\T$ as providing a distinct and statistically independent channel
to tunnel; it then corresponds to the discussion of Ref.~\cite{yoram1}
but with an infinite number of distinctly weighted open channels. We
could even imagine adding a weak magnetic field so as to interpolate
between the GOE and GUE limits as in \cite{yoram2} although we refrain
from that here. As mentioned, we have $\langle y\rangle=1$ by
construction; the second moment is
\begin{equation} \label{sndmom}
\langle y^2\rangle = 1 + {2\over\beta} 
{1-\lambda\over 1+\lambda}.
\end{equation}
The channel interpretation helps in a qualitative understanding of
this distribution as we vary $\lambda$. For small $\lambda$, only the
first $(k=0)$ channel plays any significant role and the distribution
is of the Porter-Thomas form: $\exp(-y/2)/\sqrt{2\pi y}$ and
$\exp(-y)$ for $\beta=1$ and $2$ respectively. This can be understood
analytically from (\ref{pofy}) by doing a branch-point/residue
analysis around the nearest singularity at $t=i\beta/2a$.  This
distribution is commonly used to model point tunnelling contacts
\cite{dot,yoram1,yoram2}. It is often a very accurate approximation
but its validity is not universally guaranteed, as we shall discuss.
For $\lambda$ close to unity, many channels contribute significantly,
the fluctuations around the mean are accordingly reduced and the
distribution approaches a Gaussian with variance $\sigma^2\approx
a/\beta$ (which becomes a delta function for small $a$).

For $\lambda>0$ and $y\le 0$ we close the contour
%
%
of (\ref{pofy})
in the lower half
plane; since the integrand has no singularities there, the result is
simply zero. This is consistent with the fact that $\T$ is a positive
definite operator so that Eq.(\ref{yn}) does not admit the possibility
of negative $y$.  By the same argument, any derivative of $p(y)$ is
also zero for $y\le 0$ implying a nonanalyticity at $y=0$ with $p(y)$
going to zero faster than any power of $y$ for $y$ small and positive.
In the opposite limit $y\gg\lambda$ we can expand around the first
singularity to obtain
\bea \label{asymp}
P(y;\lambda)_{\rm{GOE}} & \approx &
{\exp(-y/2a)\over \sqrt{2\pi a y}} \nonumber\\
P(y;\lambda)_{\rm{GUE}} & \approx & {\exp(-y/a) \over a}.
\eea
This falls off exponentially with $y$, and not with a power law as
observed in the chaos-assisted regime \cite{pofy_cat}.

Equations (\ref{yn}), (\ref{pofy}) and (\ref{sndmom}) remain valid
when $\lambda$ is negative. This situation arises when we compute
splittings in symmetric double wells for which the symmetry is
inversion through a point rather than reflection through an axis
\cite{us3}. In this situation $\T$ is no longer positive definite and
we admit the possibility of negative splittings (for which the odd
member of a doublet has a lower energy than the even member).  The
distribution then allows all values of $y$, positive and negative. It
decays exponentially for $|y|\gg\lambda$ as in Eq.~(\ref{asymp}) but
with different exponents for $y>0$ and $y<0$ because on doing the
integral (\ref{pofy}) we must switch from closing the integration
contour in the lower half plane to the upper half plane as $y$ changes
sign. For a given state, we can typically induce a zero splitting by
tuning one system parameter. A zero splitting means that we can
construct states which remain localised in either well for all time as 
in the one dimensional time-dependent system considered in \cite{Gross}.

At this point we contrast our results with standard random matrix
theory modelling. In section VII.H of their extensive review
\cite{brody}, Brody et al. show under rather general assumptions that
one expects a Gaussian distribution for the expectation values of an
arbitrary operator by showing that all of the moments of the
distribution approach those of a Gaussian. This can be understood as a
sort of central limit theorem. One of their assumptions is that the
operator is non-singular, {\it i.e.} does not have many zero
eigenvalues. Because of the exponential decay of the eigenvalues of
the tunnelling operator $\T$, it is effectively singular. Therefore
their conclusions do not apply to our situation and we have
non-Gaussian distributions. It is interesting to note that in the
limit $|\lambda|\rightarrow 1$, the operator $\T$ has an ever
increasing number of significant eigenvalues and the distributions do
in fact approach Gaussians, in conformity with their general
considerations.

\begin{figure}[h]
\vspace*{0.0in}
\hspace*{0.0in}\psfig{figure=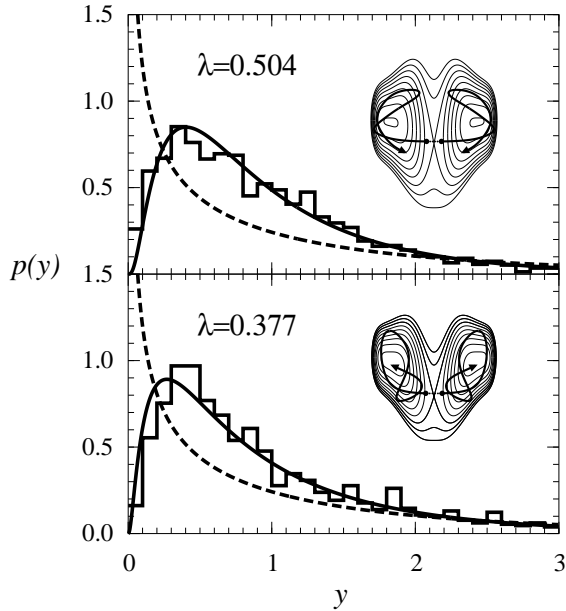,height=3.3in}
\vspace*{0.0in}
\caption
{\small Results for two typical potentials, with
$(\mu,\nu,\sigma)=(0.15,0.17,0.00)$ above and $(0.25,0.50,0.00)$
below. In both cases the energy is lower than saddle maximum by an
amount $0.1$. The continuous curves are the theoretical distributions
calculated using the appropriate values of $\lambda$ (shown). The
dashed curves show the Porter-Thomas distribution for comparison. The
insets show the corresponding instanton orbits and their real
extensions.}
\label{tisworkin}
\end{figure}

Since it is a simpler numerical task to calculate many splittings in a
double well than to calculate many resonance widths in a metastable
well, we use the former to test our predictions and note that any
conclusions apply identically to the latter. Consider the potential
\be \label{pot}
V(x,y) = (x^2-1)^4 + x^2y^2 + \mu y^2 + \nu y + \sigma x^2y.
\ee
There is a reflection symmetry in $x$ and if the energy is less than
$1-\nu^2/4\mu$ the motion is classically confined either to $x<0$ or
to $x>0$.  It is convenient to work at fixed energy in order to keep
$\lambda$ constant and we do this by quantising $q=1/\hbar$
\cite{us3}, that is, by finding those values of $\hbar$ which are
consistent with a specified choice of parameters and energy. This is
effectively what happens, for example, in scaling problems such as a
hydrogen atom in a magnetic and electric field \cite{uzer}. In
Fig.~\ref{tisworkin} we show histograms constructed from the
$q$-spectra for two choices of parameters such that the classical
dynamics is almost fully chaotic. We also show the distribution
(\ref{pofy}) with $\beta=1$ and using the corresponding values of
$\lambda$. Clearly, the numerically computed histograms are well
captured by the theoretical distribution. We show, for comparison, the
Porter-Thomas distribution which clearly fails to correctly model the
numerical data. We remark that this sort of agreement was observed for
most parameter values as long as the dynamics was fully chaotic.

\begin{figure}[h]
\vspace*{0.0in}
\hspace*{0.0in}\psfig{figure=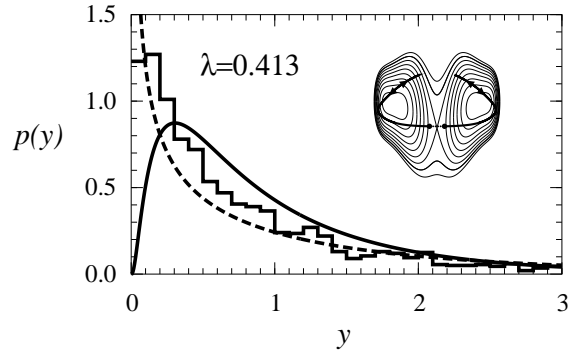,height=3.3in}
\vspace*{-1.35in}
\caption
{\small As in Fig.~1 but for parameter values
$(\mu,\nu,\sigma)=(0.25,0.40,0.254)$ for which the real extension of 
the instanton orbit is periodic and the resulting distribution is
significantly different from the RMT prediction.}
\label{scarred}
\end{figure}

In Fig.~\ref{scarred} we show an exception to the general
agreement. In this case the numerical histogram is intermediate
between the theoretical distribution (\ref{pofy}) and the
Porter-Thomas form. We attribute this to the effects of scarring
\cite{scar,kaplan}, as follows. The instanton has real turning points
where the momentum vanishes and the position is real. At these points
we can elect either to integrate in imaginary time in which case the
instanton retraces itself or to integrate in real time, in which case
we get a real trajectory. We refer to this real trajectory as the real
extension of the instanton. There is no reason why this real extension
should itself be periodic.  Typically it is not. However, the
parameters of Fig.~\ref{scarred} have been tuned so that the real
extension is in fact periodic. We find in this case that the overlaps
$x_k=\braket{k}{n}$ are no longer distributed according to the
Gaussian $P(x_k)=\exp(-x_k^2/2)/\sqrt{2\pi}$ as assumed in our
derivation --- there are relatively more large overlaps and more small
overlaps.  This effect can be explained using a recent theory of
scarring \cite{kaplan} which describes how the overlaps between a
wavepacket placed on a periodic orbit and the chaotic eigenstates
deviate from random matrix theory. In our problem the eigenvectors
$\ket{k}$ behave like wavepackets of this type when the real extension
is periodic. The effect of this deviation from random matrix theory is
to give more large splittings and more small splittings than
(\ref{pofy}) predicts and, therefore, to push the distribution in the
direction of the Porter-Thomas form. We remark that for
$\nu=\sigma=0$, the real extension is always a periodic orbit
\cite{us1us2} and we see anomalous statistics for this situation as
well.

\begin{figure}[h]
\vspace*{0cm}
\hspace*{0.0cm}\psfig{figure=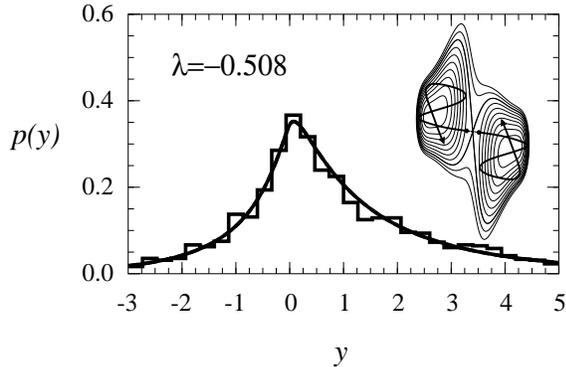,height=3.3in}
\vspace*{-1.35in}
\caption
{\small Results for the inversion-symmetric potential with
$(\mu,\tau)=(0.1,1.0)$ for which zero and negative splittings are
allowed.}
\label{xy}
\end{figure}

The final case we discuss is if the term $\nu y + \sigma x^2y$ in
(\ref{pot}) is replaced by $\tau xy$. Now the potential is symmetric
under $(x,y)\rightarrow (-x,-y)$ rather than under $(x,y)\rightarrow
(-x,y)$ (this symmetry would persist if we were to add a uniform
magnetic field). In this case $\lambda<0$ \cite{us3} and negative
splittings can occur. We present the results for a typical case in
Fig.~\ref{xy}. Again, the theoretical distribution agrees with the
histogram.

Our results are relevant to situations in which particles tunnel
out of or between chaotic regions separated by an energetic barrier.
Applications include hydrogen atoms in parallel electric and magnetic
fields where the competition between the imposed fields and the
Coulomb force causes chaotic motion while the presence of the electric
field causes tunnelling \cite{uzer}. Dissociative decays of excited
nuclei and molecules may also fall into this regime.

Another application is to conductances of quantum dots. In the
Coulomb blockade regime electrons must tunnel into and out of dots
which are thought to be chaotic. Such experiments have been done
\cite{dot_ex} leading to results which are consistent with the
Porter-Thomas distribution for the tunnelling widths, just as for
the neutron and proton resonance widths. We contend that the reason
is that the instanton path in all cases is very unstable, leading to a
small value of $\lambda$. For energies near the saddle $\lambda
\approx \exp(-2\pi\omega_y/\omega_x)$ \cite{bb} where $\omega_x$ and
$\omega_y$ are the curvatures of the potential saddle along and
transverse to the instanton, respectively. This is often small, but by
making the saddle flat in the transverse direction or sharp in the
instanton direction, it is possible to have $\lambda$ be of order
unity. It is an interesting question whether this can be arranged for
the quantum dots. One feature which helps in this regard is that we
predict a distribution which vanishes for small $y$ whereas the
Porter-Thomas distribution diverges as $1/\sqrt{y}$.  This difference
could be discernible even for rather small values of $\lambda$.

Another possibility for nonuniversal statistics would be a situation
analogous to Fig.~\ref{scarred}, where the tunnelling route is
directly connected to a real periodic orbit. This geometry could be
engineered into quantum dots and is present in the hydrogen atom
problem \cite{uzer}.
%
%
In this case we predict deviation from
the predictions of random matrix theory.
This would be similar in spirit to the recent work of
Narimanov et al. \cite{tbn} who look for dynamical effects in the
correlations of conductance peaks of quantum dots.

\end{document}